\documentclass[a4paper,reqno]{amsart}
\usepackage[foot]{amsaddr}
\usepackage{fullpage}
\usepackage{graphicx}
\usepackage{color}
\usepackage{amsthm,amssymb}
\usepackage{xfrac}

\usepackage{tikz}
\usepackage{tikz-cd}
\usepackage[hidelinks]{hyperref}
\usepackage{mathtools}
\usepackage{xcolor}
\usepackage{algpseudocode}
\usepackage{algorithm}
\usepackage{enumerate}
\usepackage{nicematrix}
\usepackage{adjustbox}

\usepackage[colorinlistoftodos,prependcaption,textsize=tiny]{todonotes}

\newcommand {\mm}[1]   {\ifmmode{#1}\else{\mbox{\(#1\)}}\fi}

\newcommand {\ceiling}[1] {{\left\lceil #1 \right\rceil}}

\newcommand{\Rspace}        {\mm{{\mathbb R}}}

\newcommand{\Zspace}        {\mm{{\mathbb Z}}}
\newcommand{\Hgroup}[1]     {\mm{\sf H}_{#1}}
\newcommand{\Betti}[1]      {\mm{\beta}_{#1}}

\newcommand{\Delaunay}[1]   {\mm{{\rm Del}{({#1})}}}
\newcommand{\Alpha}[2]      {\mm{{\rm Alf}_{#1}{({#2})}}}

\newcommand{\Dgm}[2]        {\mm{{\rm Dgm}_{#1}{({#2})}}}

\newcommand{\Radiusf}       {\mm{{g}}}

\newcommand{\Kfun}          {\mm{{f_K}}}
\newcommand{\Lfun}          {\mm{{f_L}}}

\newcommand{\KLfun}         {\mm{{f_{K,L}}}}

\newcommand{\rank}[1]       {\mm{{\rm rank\,}{#1}}}

\newcommand{\card}[1]       {\mm{{\#}{#1}}}

\newcommand{\conv}[1]       {\mm{\rm conv}{({#1})}}

\newcommand{\image}[2]      {\mm{\rm img}_{#1}{{\,}#2}}
\newcommand{\kernel}[2]     {\mm{\rm ker}_{#1}{\,#2}}
\newcommand{\coker}[2]      {\mm{\rm cok}_{#1}{\,#2}}

\newcommand{\norm}[1]       {\mm{\|{#1}\|}}

\newcommand{\Skip}[1]       {}

\definecolor{blue-red}{rgb}{0.8, 0.00, 0.95}

\newcommand{\Length}[1]     {\mm{|{#1}|}}
\newcommand{\MST}[1]        {\mm{{\rm MST}{({#1})}}}

\newcommand{\ee}            {\mm{\varepsilon}}

\theoremstyle{definition}

\newtheorem{theorem}{Theorem}
\numberwithin{theorem}{section}

\newtheorem{lemma}[theorem]{Lemma}

\newtheorem{definition}[theorem]{Definition}

\numberwithin{equation}{section}

\title{Chromatic Topological Data Analysis}

\author{Sebastiano Cultrera di Montesano$^1$}
\email{$^1$sebastiano.cultrera@ist.ac.at}
\author{Ond\v{r}ej Draganov$^1$}
\email{$^2$ondrej.draganov@ist.ac.at}
\author{Herbert Edelsbrunner$^1$}
\email{$^3$edels@ist.ac.at}
\author{Morteza Saghafian$^1$}
\email{$^4$morteza.saghafian@ist.ac.at}

\address{$^{1}$ISTA (Institute of Science and Technology Austria), Kloster\-neu\-burg, Austria}


\begin{document}

\begin{abstract}
  Exploring the \emph{shape} of point configurations has been a key driver in the evolution of TDA (short for \emph{topological data analysis}) since its infancy. 
  This survey illustrates the recent efforts to broaden these ideas to model spatial interactions among multiple configurations, each distinguished by a color.
  It describes advances in this area and prepares the ground for further exploration by mentioning unresolved questions and promising research avenues while focusing on the overlap with discrete geometry.
\end{abstract}

\maketitle

\section{Introduction}
\label{sec:1}

Discrete point sets are fundamental objects in discrete geometry, captivating its community for ages.  
In today's data-driven context, information frequently emerges as a point cloud, and understanding its shape is often a critical step toward unraveling its essence. 
An important step in this direction was taken in the 80s and 90s of last century, with the theory of alpha complexes introduced by Edelsbrunner and collaborators \cite{EdKiSe83, EdMu94} which was designed to quantitatively describe a spatial configuration of points.
The core idea behind this mathematical construction involves expanding discs around each data point with progressively larger radii and monitoring the evolution of the homological features (i.e., the connected components and the cycles) as the radius expands. 
Persistent homology, which emerged around the turn of the century, pushed this concept further by allowing not only to identify radii at which homological features appear and disappear, but also pair these events to quantify how long each feature persists. 
The persistence of each homological feature is compactly stored in a bookkeeping device referred to as the the persistent diagram \cite{EdHa10}. 
This methodology is now recognized as a cornerstone in the mathematical field of topological data analysis (TDA), providing a foundational framework for the discipline.

\smallskip
Can similar ideas be adapted to scenarios involving multiple point sets to closely examine their spatial interactions?
This question arises for example in material science, but also in biology, where it is spurred by the desire to describe the interplay between the tumor and the immune environment, known as the tumor immune microenvironment. 
Indeed, there is a growing body of evidence suggesting that the spatial configuration of different cell types can significantly influence disease progression and patient response to treatment across various cancer types \cite{PFRT22}.
This perspective drove the exploration of how to broaden the theory of alpha complexes to encompass chromatic point sets, where each point is attributed a color. 
The goal is to engineer stable and multi-scale topological descriptors, underpinned by suitable discrete structures to facilitate their computation, thereby enhancing our ability to quantitatively describe the complex interplay within chromatic point sets effectively.

This endeavor opens the door to novel combinatorial inquiries in discrete geometry, which we will describe in Section \ref{sec:3} and \ref{sec:5}, as well as algebraic investigations detailed in Section \ref{sec:4}. 
We aim to provide a comprehensive summary of the current landscape, delineating existing knowledge while also proposing future research avenues and posing unresolved questions. As TDA progresses, we anticipate that it will reinvigorate longstanding queries in discrete and computational geometry, potentially setting a constructive path for further exploration and discovery.

\smallskip
While this survey focuses on chromatic variants of the alpha complexes, we acknowledge that there are other discrete structures amenable to a topological study of chromatic point sets. 
For example, the developments described in this survey could be based on the \v{C}ech rather than the alpha complex, with almost no differences, except that the complexes would be significantly larger, making computational experiments harder to perform in practice.
Recently, a topological descriptor known as a mixup barcode, characterizing interactions between two point clouds using Vietoris--Rips complexes, was introduced \cite{WAWB24}.
Additionally, Dowker complexes and witness complexes have been suggested as possible candidates for encoding spatial relations in the tumor microenvironment \cite{Sto23}. 
A shortcoming of these complexes is their lack of stability.
Moreover, the Dowker complex is limited to two interacting point sets, while the witness complex requires a choice of `landmark points', and it is unclear how to choose these in practice.
Finally, the comparison of colors can also be modeled as a multi-parameter persistent module.
The representation theoretic questions of decomposing such a module are significantly more difficult than in the $1$-parameter case.
We mention recent work on the decomposition into so-called \emph{interval indecomposables} \cite{AENY23}, which is computationally more feasible and may have connections to the work described in this survey.

\smallskip \noindent \textbf{Outline.}
Section~\ref{sec:2} reviews two-dimensional alpha complexes and their persistent homology.
Section~\ref{sec:3} extends these constructions to the chromatic case. 
Section~\ref{sec:4} focuses on the proposed topological descriptors for chromatic point sets, referred to as the \emph{6-pack of persistent diagrams}.
Section~\ref{sec:5} aims at extracting \emph{mingling numbers} from the 6-pack and highlights connections to discrete geometry.
Section~\ref{sec:6} concludes the paper.


\section{TDA in a Nutshell}
\label{sec:2}
We present the simplest scenario of the TDA pipeline focusing on the case of $2$-dimensional alpha complexes. 
This section sets the stage for what comes next.
For a more comprehensive introduction to topological data analysis, we recommend \cite{Car09,EdHa10}.

\subsection{Alpha Complexes}
\label{sec:2.1}

\begin{figure}[htp]
    \centering
    \includegraphics[width=0.3\textwidth]{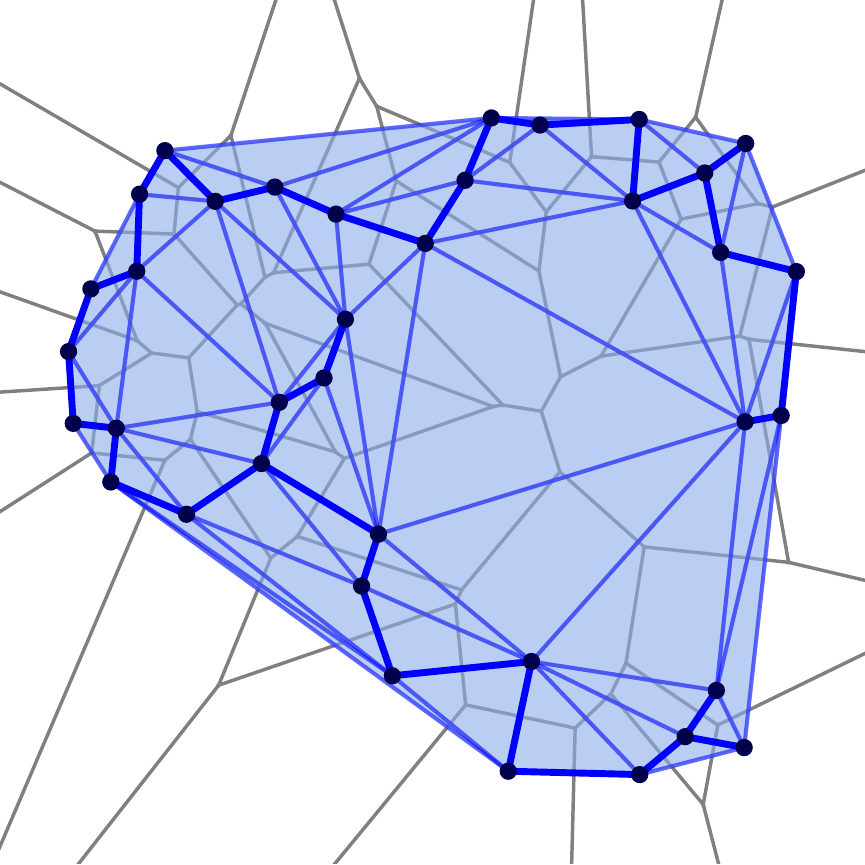}
    \hspace{5mm}
    \includegraphics[width=0.3\textwidth]{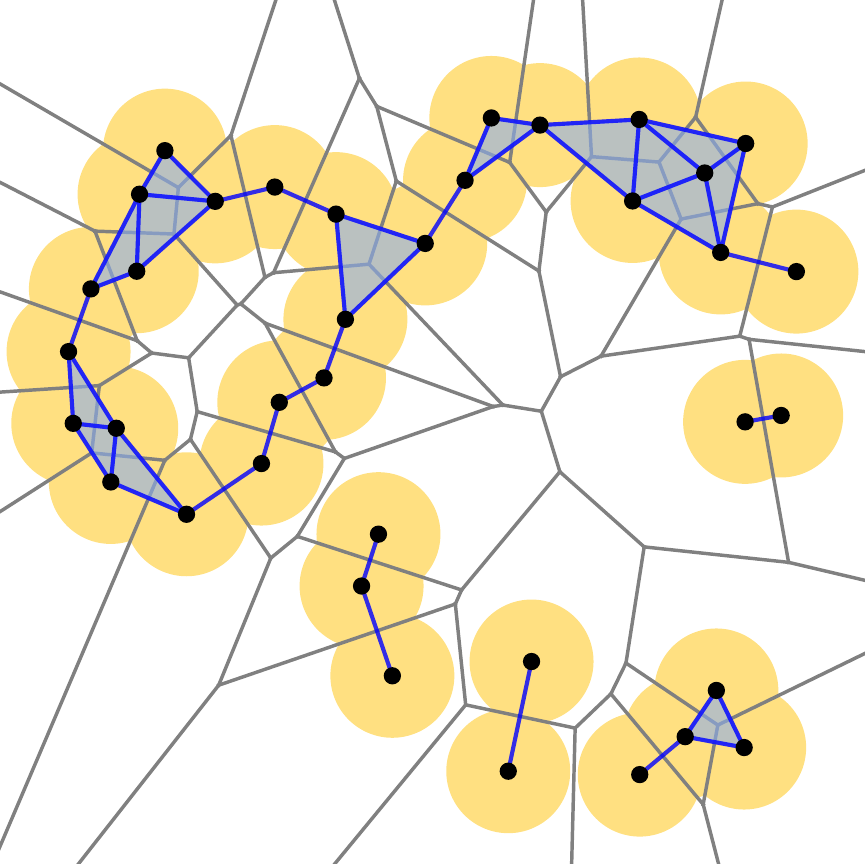}
    \hspace{5mm}
    \includegraphics[width=0.3\textwidth]{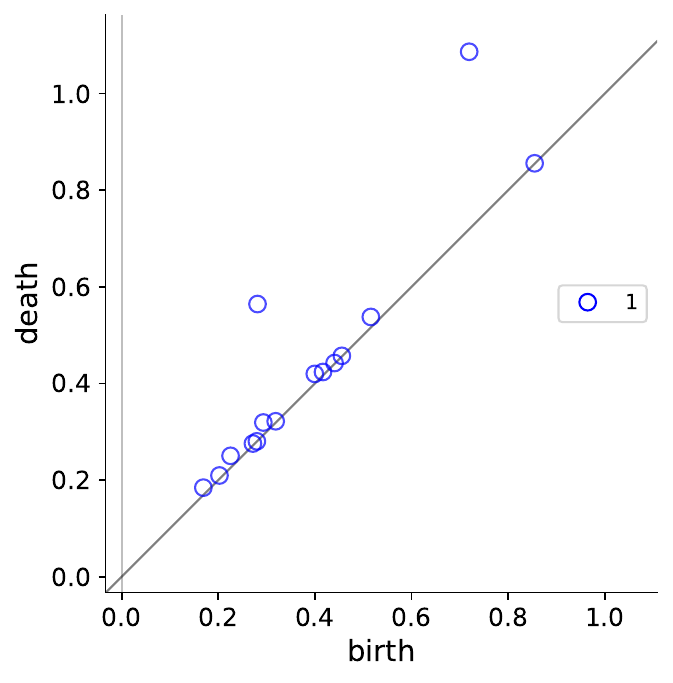}
    \caption{\footnotesize \emph{Left}: the \emph{shaded} Delaunay mosaic superimposed on the Voronoi tessellation of a finite set of points.
    The \emph{thick blue} edges of the mosaic form the (Euclidean) minimum spanning tree of the points.
    \emph{Middle}: the alpha complex is dual to the decomposition of the union of disks into convex regions by the Voronoi tessellation.
    \emph{Right}: the corresponding 1-dimensional persistence diagram tracs the birth and death of the loops. 
    Note in particular the two dots far above the diagonal, which represent the two large circles suggested by the points in the set.}
    \label{fig:two_circles}
\end{figure}

Let $A$ be a finite set of points in Euclidean plane, $\Rspace^2$, and let $r \geq 0$.
The most direct definition of the alpha complex of $A$ and $r$ \cite{EdHa10} starts with the \emph{Voronoi tessellation} \cite{Vor070809} and its dual, the \emph{Delaunay mosaic} \cite{Del34}.
This tessellation decomposes the plane by drawing the convex polygons of points closest to every $b \in A$.
We streamline the discussion by taking liberties in assuming the points are in \emph{general position}, which includes that no four lie on a common circle.
For such points, at most three polygons may share a vertex, and any such triplet defines a triangle in the dual mosaic.
While the Voronoi tessellation is a collection of polygons, the Delaunay mosaic is a complex consisting of vertices (the points in $A$), edges (connecting points generating neighboring polygons), and triangles (filling the area between their three edges), denoted $\Delaunay{A}$; see the left panel in Figure~\ref{fig:two_circles}.

\smallskip
Write $A_r$ for the points $x \in \Rspace^2$ at distance at most $r$ from at least one point in $A$, and note that the Voronoi tessellation decomposes $A_r$ into convex sets, each the intersection of a disk with a convex polygon. 
Taking the dual of this decomposition, we get the \emph{alpha complex} for radius $r$, denoted $\Alpha{r}{A}$, which we observe is a subcomplex of the Delaunay mosaic; see the middle panel in Figure~\ref{fig:two_circles}. 
$\Alpha{r}{A}$ is a simplicial complex for every $r$, and when $r$ increases, $\Alpha{r}{A}$ gains a possibly empty collection of simplices (vertices, edges, and triangles).
For each simplex there is a threshold beyond which it belongs to the alpha complex.
Specifically, this threshold is the radius of the smallest \emph{empty circumcircle}; that is: the smallest circle that passes through the vertices of the simplex and does not enclose any of the points in $A$.
Write $f \colon \Delaunay{A} \to \Rspace$ for the function that maps each simplex to this threshold, refer to $f$ as the \emph{radius function} on the Delaunay mosaic, and observe that the alpha complexes are its sublevel sets: $\Alpha{r}{A} = f^{-1} [0, r ]$ for every $r$.

\subsection{Persistent Homology}
\label{sec:2.2}

When we vary the radius, the alpha complex changes its shape, and we use the algebraic formalism of homology groups to capture shape topologically.
In the plane, this just means that we keep track of connected components and cycles.
We sidestep many of the technicalities by restricting the discussion to the simplest of settings.
In particular, we assume a complex in two rather than $d$ dimensions, we consider homology for a finite complex rather than more general sets, and we use modulo-$2$ arithmetic so that formal sums simplify to sets, and addition simplifies to taking the symmetric difference.
We refer to texts in algebraic topology for the general theory \cite{Hat02}.

\smallskip
For a complex in $\Rspace^2$, there are only two possibly non-trivial homology groups, one for dimension $0$ and the other for dimension $1$.
\Skip{
  The group in dimension $0$ is made up of connected components, each represented by one of its vertices.
  A \emph{class} in $\Hgroup{0}$ is a collection of such components and consists of all possible representations.
  We say any two such representations are \emph{homologous} as their points can be paired up so that corresponding points belong to the same component.
  The \emph{sum} of two such collections is the symmetric difference, which is again a collection of components.]]
} 
The group in dimension~$0$ is the vector space over $\Zspace / 2 \Zspace$ generated by the connected components. 
Each vector is a collection of components, and the sum of two is the symmetric difference. 
In the formal definition of $\Hgroup{0}$, a vector is a \emph{class} of homologous sets of vertices, in which two sets are \emph{homologous} if their symmetric difference intersects each component in an even number of vertices. 
The sum of two classes is then the class of the symmetric difference of any representatives.
Moving up one dimension, $\Hgroup{1}$ captures cycles in the complex.
Formally, a \emph{cycle} is a set of edges such that each vertex belongs to an even (and possibly zero) number of these edges.
A cycle thus looks like a closed curve, but it may not be connected, and there may be crossings.
For example, the three edges of a triangle form a cycle, and if the triangle belongs to the complex, then this cycle is \emph{homologous} to the empty cycle, or \emph{trivial}.
More generally, two cycles are \emph{homologous} if they add up to a sum of trivial cycles.
A \emph{class} in $\Hgroup{1}$ is then a set of homologous cycles, and the sum of two classes is the class represented by the symmetric difference of any two representatives.
Since $\Hgroup{0}$ and $\Hgroup{1}$ are vector spaces, we may consider their ranks, which are referred to as the \emph{Betti numbers} of the complex, $\Betti{0} = \rank{\Hgroup{0}}$ and $\Betti{1} = \rank{\Hgroup{1}}$.
We think of them as the number of connected components and the number of holes (or cycles) in the complex.

\smallskip
To make the step from homology to persistent homology, we return to the sequence of alpha complexes and write $r_0 < r_1 < \ldots < r_n$ for the radii of the simplices in $\Delaunay{A}$, and $K_i = f^{-1} [0,r_i]$ for the $i$-th alpha complex.
Persistence arises from considering the homology groups of the $K_i$ and relations between these groups.
To describe this with an example, suppose the difference between $K_i$ and $K_{i+1}$ is a single simplex, whose dimension is $p$.
Adding this $p$-simplex to $K_i$ either increases the rank of $\Hgroup{p}$ by one, or it decreases the rank of $\Hgroup{p-1}$ by one.
In the former case, we say the $p$-simplex gives \emph{birth} to a $p$-dimensional homology class, and in the latter case, we say it gives \emph{death} to a $(p-1)$-dimensional homology class.
For example, every vertex gives birth to a connected component, every edge either gives birth to a cycle (if it connects two vertices in the same component) or death to a component (if it connects two different components), and in $\Rspace^2$ every triangle gives death to a cycle.

\smallskip
The main insight of persistence is the existence of a canonical injection from the deaths to the births that encodes a rich amount of information about the complexes in the sequence.
To explain, let $i \leq j$ and note that the inclusion $K_i \subseteq K_j$ induces a (linear) map from $\Hgroup{p} (K_i)$ to $\Hgroup{p} (K_j)$ for each $p$.
We call the image of $\Hgroup{p} (K_i)$ in $\Hgroup{p} (K_j)$ a \emph{persistent homology group} and its rank a \emph{persistent Betti number}.
This number counts the $p$-dimensional homology classes born before or at $r_i$ that did not yet die at $r_j$.

\smallskip
The birth of a class at $r_i$ that dies at $r_j$ is sometimes drawn as the half-open interval $[r_i, r_j)$ and at other times as the point $(r_i, r_j)$.
The multi-set of such intervals is the \emph{persistence barcode}; see the panels on the right in Figure~\ref{fig:monkey-set}, while the multi-set of points is the \emph{persistence diagram}; see the right panel in Figure~\ref{fig:two_circles}.
The two graphical representations of persistence have their own advantages and disadvantages.
The barcode represents the rank of the homology groups of $K_i$ by the number of intervals that contain $r_i$, which is more intuitive than the number of points in the upper-left quadrant anchored at the point $(r_i, r_i)$ in $\Rspace^2$, which is the representation of the rank by the persistence diagram.
On the other hand, the stability of persistence is easier to state for the points.
It is formulated in terms of the \emph{bottleneck distance}, which is the $L_\infty$-length of the longest edge in a minimizing perfect matching between two persistence diagrams, where we borrow points from the diagonal whenever this shortens the distance.
Given two sets of points, $A, B \subseteq \Rspace^2$, the stability theorem originally proved in \cite{CEH07} implies that the bottleneck distance between the persistence diagrams of the radius functions on $\Delaunay{A}$ and $\Delaunay{B}$ is bounded from above by the Hausdorff distance between $A$ and $B$.

\subsection{Minimum Spanning Trees}
\label{sec:2.3}

A \emph{tree} is a connected graph without cycle, and it \emph{spans} a connected graph if it is a subgraph that touches all vertices.
In the Euclidean setting, the vertices are the points in $A \subseteq \Rspace^2$, and we consider all spanning trees of the complete graph on these points.
The \emph{length} of an edge is the Euclidean distance between its endpoints, and the \emph{length} of a tree is the sum of edges lengths.
A classic topic in discrete mathematics is the \emph{minimum spanning tree}, or \emph{MST} for short, which minimizes length.
Our motivation for discussing its construction is the close relation to the $0$-dimensional persistent homology of the radius function on $\Delaunay{A}$.
There is a second motivation, which is the application of minimum spanning trees to measuring the mingling of point sets, as studied in Section~\ref{sec:5} of this survey.

\smallskip
The spanning trees of a connected graph form a matroid, which essentially means that any two spanning trees have an edge each, such that exchanging these edges produces two new spanning trees.
Of course, if the two trees differ by only these two edges, then exchanging them turns the trees into each other.
By exploiting this property, it is possible to construct the MST greedily, by adding one edge at a time.
Most convenient for our purposes is Kruskal's algorithm \cite{Kru56}, which starts by sorting the edges in the order of increasing length.
Thereafter, it adds the shortest remaining edge to the evolving tree, unless this edge forms a cycle together with the previously added edges, in which the edge is discarded.
Then the edge is removed from the list, and the process repeats until the tree is complete and spans all $n$ points.
This happens when the tree consists of $n-1$ edges.

To relate the MST to the $0$-dimensional persistence diagram, we note that the edges added to form the tree are exactly the ones that give death in $0$-dimensional homology.
A death happens when the radius reaches half the length of the responsible edge.
The sum of death values is thus half the length of the MST.
Since all components are born at $r_0 = 0$, this is also the total length of the intervals in the $0$-dimensional persistence barcode or, equivalently, the $1$-norm of the $0$-dimensional persistence diagram; see Sections~\ref{sec:4} and \ref{sec:5}.
All these edges belong to $\Delaunay{A}$, which implies that in the Euclidean setting the MST is necessarily a subgraph of the Delaunay mosaic.

\smallskip
It is instructive to observe that an edge that connects two vertices in a tree but does not belong to the tree forms a unique cycle with a subset of the tree edges.
This cycle is but one representative of the cycle that is born when we add this edge to the complex.
See Figure~\ref{fig:two_circles} for a concrete example.
The alpha complex shown in the middle panel has only one hole, and a natural choice of cycle that goes around this hole is the one that encloses the minimum area.
There are however other, homologous cycles, and there is a unique one obtained by adding a shortest remaining edge rejected by Kruskal's algorithm to the MST shown in the left panel.

\section{Discrete Chromatic Structures}
\label{sec:3}

In this section, we extend the core notions introduced above to the setting where each point is endowed with a color.
We begin with the chromatic extension of the Delaunay mosaic (which generalizes the \emph{coupled alpha shape} in \cite{ReBo23}) and review what is known about its size and sublevel sets.

\subsection{Chromatic Delaunay Mosaic}
\label{sec:3.1}

Letting $s+1$ be the number of colors, we use an extra dimension for each color but the first.
Specifically, if the points are in $\Rspace^d$, we map them to $\Rspace^{s+d}$ and construct the chromatic extension of the Delaunay mosaic as the ordinary Delaunay mosaic in $\Rspace^{s+d}$.
While this may seem excessive, we will see shortly that in many cases, the size of the chromatic extension is barely larger than that of the Delaunay mosaic of the original points in $\Rspace^d$.
To explain the construction, we let $A$ be a finite set in $\Rspace^d$, $\sigma = \{0,1,\ldots,s\}$ a collection of colors, $\chi \colon A \to \sigma$ a coloring, and $A_j = \chi^{-1} (j)$ the points with color $j$.
Recall that the \emph{standard $s$-simplex} is the convex hull of the $s+1$ unit coordinate vectors in $\Rspace^{s+1}$.
To map this simplex to $s$ dimensions, we identify $\Rspace^s$ with the $s$-plane defined by $x_1 + x_2 + \ldots + x_{s+1} = 1$ in $\Rspace^{s+1}$ and parametrize it with the inherited $s+1$ barycentric coordinates.
We write $\Rspace^{s+d} = \Rspace^s \times \Rspace^d$, implying the explicit embeddings of $\Rspace^s$ and $\Rspace^d$ in $\Rspace^{s+d}$.
\begin{definition} 
  \label{dfn:chromatic_Delaunay_mosaic}
  Writing $u_0, u_1, \ldots, u_s$ for the vertices of the standard $s$-simplex in $\Rspace^s$, we set $A_j' = u_j + A_j \subseteq u_j + \Rspace^d$, for each $0 \leq j \leq s$, and $A' = A_0' \cup A_1' \cup \ldots \cup A_s'$.
  The \emph{chromatic Delaunay mosaic} of $\chi$, denoted $\Delaunay{\chi}$, is the ordinary Delaunay mosaic of $A'$ in $\Rspace^{s+d}$.
\end{definition}
\begin{figure}[htb]
  \centering \vspace{0.0in}
  \includegraphics[width=0.4\textwidth]{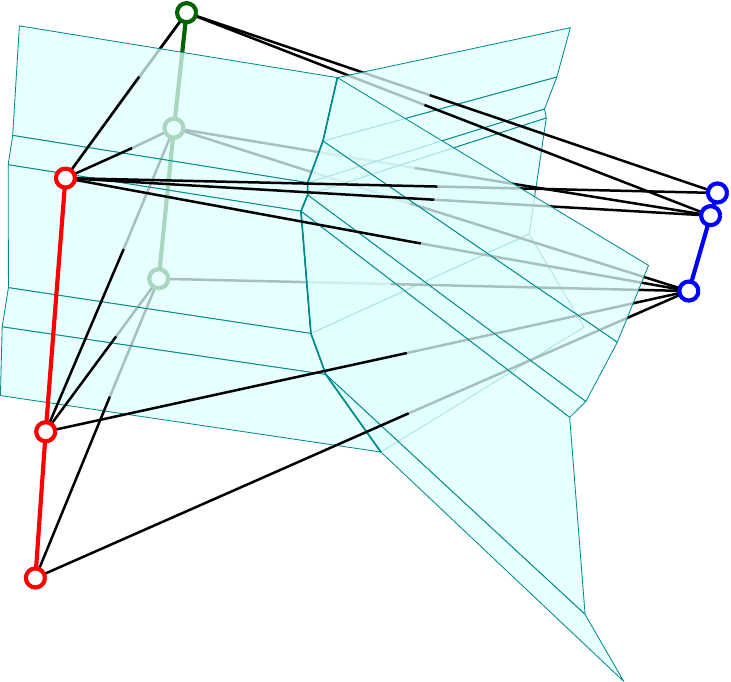}
  \caption{\footnotesize The chromatic Delaunay mosaic of three finite sets in $\Rspace^1$ together with a corresponding stratification of space.
  The points of each set are placed on a copy of $\Rspace^1$ orthogonal to the $2$-plane that carries the standard triangle.
  The stratification consists of Voronoi cells that separate the three colors by forming a $1$-dimensional stratum geometrically located between the three lines, and three $2$-dimensional strata, one between any two of the lines.}
  \label{fig:chromatic}
\end{figure}
Figure~\ref{fig:chromatic} illustrates the concept by showing the Delaunay mosaic together with the cells in the dual Voronoi tessellation that separate the different colors.
Similarly, we apply the construction to a subset of the colors, $\tau \subseteq \sigma$, and write $\Delaunay{\chi|\tau}$, in which $\chi|\tau$ is our notation for the restriction of $\chi$ to $\chi^{-1} (\tau)$.
This mosaic lives in $\Rspace^{t+d}$, in which $t = 1+\card{\tau}$.
It is not difficult to see that $\Delaunay{\chi|\tau}$ is a subcomplex of $\Delaunay{\chi}$.
This will be important in Section~\ref{sec:4}, where we study maps induced by this inclusion.
To state the property formally, we call a cell in $\Delaunay{\chi}$ \emph{$\tau$-colored} if the colors of its vertices belong to $\tau$, and observe that the $\tau$-colored cells form a subcomplex of $\Delaunay{\chi}$. 
\begin{lemma} 
  \label{lem:sub-chromatic_Delaunay_subcomplexes}
  Let $A \subseteq \Rspace^d$ be finite, $\chi \colon A \to \sigma$ a coloring, and $\tau \subseteq \sigma$.
  Then the subcomplex of $\tau$-colored simplices in $\Delaunay{\chi}$ is $\Delaunay{\chi|\tau}$.
\end{lemma}

The number of simplices in a Delaunay mosaic---and therefore also in a chromatic Delaunay mosaic---is indicative of the time it takes to compute it.
There are many algorithms to choose from; see e.g.\ \cite{BCKO08}, and as a rule of thumb the time they take is the number of simplices times a factor, which is usually somewhere between a constant and linear in the number of points.
This motivates the combinatorial question of counting the simplices that appear in a chromatic Delaunay mosaic.

\subsection{Size or Number of Simplices}
\label{sec:3.2}

Call the number of simplices in a Delaunay mosaic its \emph{size}, which depends on the number of points, denoted $n$, the dimension, $d$, the number of colors, $s$, but also on the way the points are distributed.
We review the results in \cite{BCDES22}, which assume that $d$ and $s$ are constants.
We also consider locally finite but possibly infinite sets, namely Delone sets and Poisson point processes as examples of reasonable well packed and random sets in $\Rspace^d$, respectively.
To facilitate the comparison with the finite sets, we count the simplices within a sufficiently large ball centered at the origin.

\medskip
The results are summarized in Table~\ref{tbl:sizebounds}.
In particular, we have upper bounds for three types of point sets (`worst-case', `well packed', and `random') assuming the colors are assigned at random.
For worst-case and well packed points, the bounds are asymptotically tight. 
Note the conspicuous absence of the number of colors in most bounds given, and this despite the fact that the chromatic Delaunay mosaic is an $(s+d)$-dimensional complex.
For a stationary Poisson point process with finite intensity, the expected density exists, which implies that within a sufficiently large ball, the expected size is proportional to the expected number of points.
\begin{table}[hbt]
    \footnotesize \centering
    \begin{tabular}{c||c|c|c||c}
      & \multicolumn{3}{c||}{chromatic Delaunay mosaic in $\Rspace^{s+d}$}
      & \multicolumn{1}{c}{Delaunay mosaic in $\Rspace^{s+d}$} \\
      size & worst-case points & well packed points & random points & (one color) \\ \hline \hline 
      & & & & \\ [-2mm]
      worst-case colors & $n^{\min\{d, \ceiling{{(s+d)}/{2}}\}}$ & $\min\{m^2, n^2\}$ in $\Rspace^2$ $(^*)$ & ? &  $n^{\ceiling{{(s+d)}/{2}}}$ \\
      & \cite[Section~4]{BCDES22} & \cite[Theorem~4.7]{BCDES22} & & \cite{ClSh89} \\ [1mm] \hline
      & & & & \\ [-2mm]
      random colors & $n^{\ceiling{{d}/{2}}}$ & $n$ & $n$ & \\
      & \cite[Theorem~4.2]{BCDES22} & \cite[Theorem~4.4]{BCDES22} &  \cite[Theorems~5.1, 5.2]{BCDES22} & 
      \\ [+2mm]
    \end{tabular}
    \caption{\footnotesize Asymptotic size bounds for chromatic Delaunay mosaics of $n$ points in $\Rspace^d$ with $s+1$ colors.
    Constant factors are not shown.
    For the case of a well packed set and worst-case colors, we have a result only in $\Rspace^2$ $(^*)$, 
    in which $m$ is the spread, which is at least a constant times $\sqrt{n}$.
    For comparison, we state the known maximum size of the (mono-chromatic) Delaunay mosaic of $n$ points in $\Rspace^{s+d}$ in the last column on the right \cite{ClSh89}.}
    \label{tbl:sizebounds}
\end{table}

We have partial results for worst-case assignments of the colors.
The bounds for worst-case points are straightforward and again asymptotically tight.
For well packed points, we have a result in $\Rspace^2$, proving that the size is at most quadratic in the spread (the diameter divided by the minimum interpoint distance), denoted $m$.
Hence, the size is at most $O(m^2)$, and thus $O(n)$ if $m = O(\sqrt{n})$.
This $O(m^2)$ bound is tight for all values of $m$ between a constant times $\sqrt{n}$ and $n$.
We lack bounds for well packed points and worst-case colors beyond two dimensions and for random points and worst-case colors beyond one dimension.
Besides filling the remainder of this table, there are also open-ended research directions suggested by our chromatic constructions:
\smallskip \begin{itemize}
  \item How much of a difference does the change of color of a small number of points make to the size of the chromatic Delaunay mosaic?
  \item What is the variance of the size assuming the coloring is random?
\end{itemize} \smallskip
The questions are motivated by the desire for stability with respect to coloring. Such stability is motivated by the representation of a (biological) cells as a point and its type as a color, where misclassifications can happen. 
More generally it is of interest to understand how the structure of the mosaic changes when we alter the colors of the points.

\subsection{Chromatic Alpha Complexes}
\label{sec:3.3}

\begin{figure}[htb]
    \centering
    \includegraphics[width=.38\textwidth]{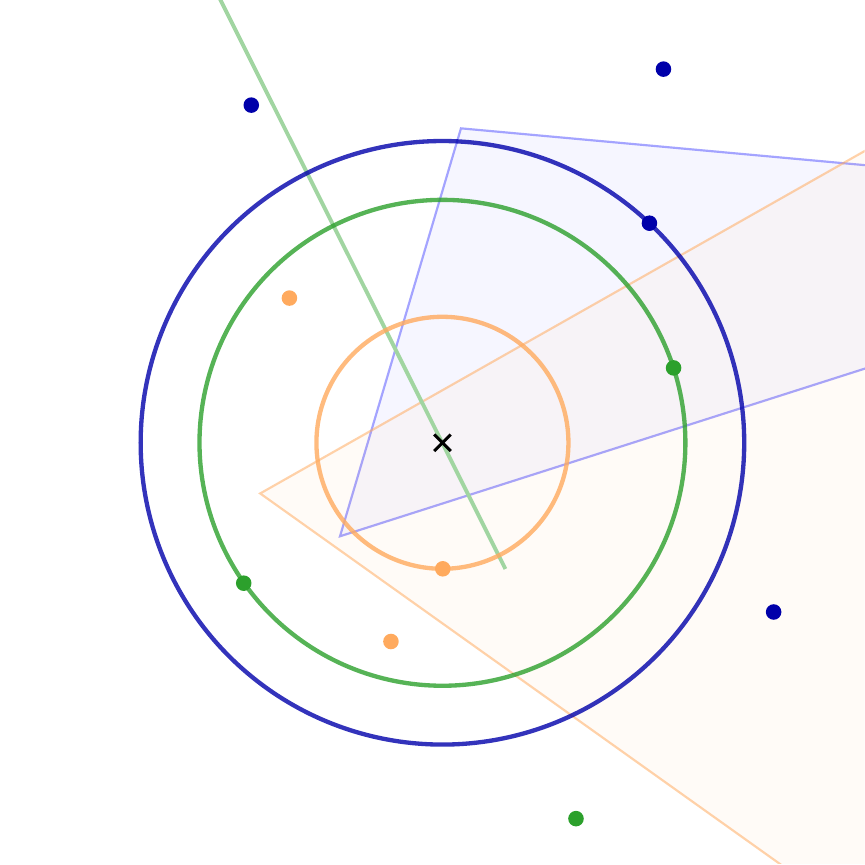} \hspace{10mm}
    \includegraphics[width=.38\textwidth]{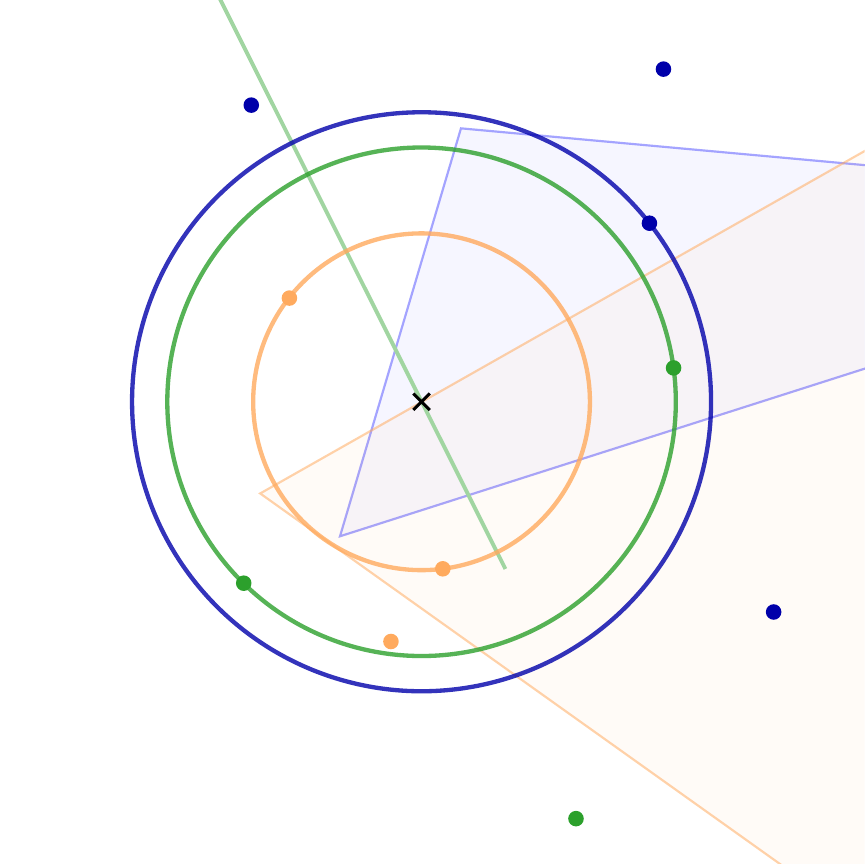}
    \caption{\footnotesize Two empty stacks in $\Rspace^2$ that pass through one blue point, two green points, and one orange point forming a simplex in $\Delaunay{\chi}$. 
    (In fact, the stack on the \emph{right} passes through \emph{two} orange points, so it also passes through the \emph{one} orange point that lies on the \emph{left} orange circle.)
    The set of centers of all empty stacks that pass through these four points is the intersection of three Voronoi cells: a blue $2$-cell, a green $1$-cell, and an orange $2$-cell.
    The \emph{right panel} shows the smallest empty stack in this collection: its center lies on the boundary of the intersection of Voronoi cells, which is the reason why one of its circles passes through an extra point.
    }
    \label{fig:stack}
\end{figure}

A direct analogy of the characterization of the Delaunay complex with empty spheres is the characterization of the chromatic Delaunay complex with what we call empty stacks. 
A \emph{$\sigma$-stack} in $\Rspace^d$ is a collection of $s+1$ concentric $(d-1)$-spheres, one for each color in $\sigma$; see Figure~\ref{fig:stack}. 
We drop $\sigma$ from the notation if it is clear from the context. 
The \emph{radius} of the stack is the maximum radius of its spheres, and its \emph{center} is the common center of the spheres. 
We label the spheres $S_j$, $j \in \sigma$, and say the stack is \emph{empty} if $S_j$ is empty of points in $A_j = \chi^{-1} (j)$, for each $j \in \sigma$.
We say the stack \emph{passes through} $\nu \subseteq A$ if $S_j$ passes through all the points of $\nu \cap A_j$, for each color $j \in \sigma$.
\begin{lemma} 
  \label{lem:characterization_with_empty_stacks}
    Let $A \subseteq \Rspace^d$ be finite, $\chi \colon A \to \sigma$ a coloring, and $\nu \subseteq A$ a collection of points. 
    Then $\conv{\nu} \in \Delaunay{\chi}$ iff there exists an empty stack of spheres that passes through $\nu$. 
\end{lemma}

Like in the mono-chromatic setting, we define chromatic alpha complexes as sublevel sets of the radius function defined on the chromatic Delaunay complex.
We recall that the radius of a stack is the radius of its largest sphere.
\begin{definition}
  \label{dfn:chromatic_alpha_complex}
    Let $\chi \colon A \to \sigma$ be a chromatic point set, and $\Radiusf \colon \Delaunay{\chi} \to \Rspace$ the \emph{radius function} defined by mapping $\conv{\nu} \in \Delaunay{\chi}$ to the radius of the smallest empty stack that passes through $\nu$.
    The \emph{chromatic alpha complex} of $\chi$ with radius $r \in \Rspace$ is $\Alpha{r}{\chi} = \Radiusf^{-1} [0,r]$.
\end{definition}

Lemma~\ref{lem:sub-chromatic_Delaunay_subcomplexes} extends to chromatic alpha complexes; that is: the restriction of $\Alpha{r}{\chi}$ to any subset of the colors is itself a chromatic alpha complex and also a subcomplex of $\Alpha{r}{\chi}$.
Furthermore, for any $r \geq 0$, $\Alpha{r}{\chi}$ has the same homotopy type as $\Alpha{r}{A}$.
A stronger result holds, which will be discussed in the next subsection.
An algorithm that computes the radius function on a chromatic Delaunay mosaic in constant time per simplex is described in \cite{CDES24}.
Similar to the algorithm for points without color \cite{EdMu94}, it proceeds in the direction of decreasing dimension and either copies the value of a coface computed earlier, or assigns the radius of the smallest empty stack.
The details in the chromatic case are however more complicated because a stack has more parameters than a single sphere.

\subsection{Generalized Discrete Morse Property}
\label{sec:3.4}

A non-trivial property of the chromatic Delaunay mosaic is that the two radius functions, $f \colon \Delaunay{A} \to \Rspace$ and $g \colon \Delaunay{\chi} \to \Rspace$, are both generalized discrete Morse and have the same critical simplices.
Among other things, this implies that $\Alpha{r}{A}$ and $\Alpha{r}{\chi}$ have the same homotopy type for every $r$.
To explain what the mentioned property of $f$ and $g$ means, we recall that the \emph{interval} defined by two simplices, $P \subseteq R \in \Delaunay{A}$, is the collection of simplices, $Q \in \Delaunay{A}$, that satisfy $P \subseteq Q \subseteq R$.
The preimages of the values of $f$ partition $\Delaunay{A}$ into subsets of simplices with shared value, and if all these subsets are disjoint unions of maximal intervals, then $f$ is \emph{generalized discrete Morse}.
For comparison, if all these subsets are singletons ($P = R$) or pairs (either $Q = P$ or $Q = R$ for every $Q$ in the interval), then $f$ is \emph{discrete Morse}, as defined by Forman~\cite{For98}.
The fact that $f$ is generalized discrete Morse has been folklore and was explicitly proved in \cite{BaEd17}.
The proof that $g$ has this property as well is more recent and can be found in \cite[Theorem~4.6]{CDES22}.
\begin{theorem} 
  \label{thm:generalized_discrete_Morse}
  Let $A$ be a finite set of points in general position in $\Rspace^d$, and $\chi \colon A \to \sigma$ a coloring.
  Then the chromatic radius function on the chromatic Delaunay mosaic, $\Radiusf \colon \Delaunay{\chi} \to \Rspace$, is generalized discrete Morse.
\end{theorem}

A simplex is \emph{critical} if it is the only simplex in its interval ($P = Q = R$).
All other simplices are \emph{non-critical}.
If the difference between a sublevel set and the next is a critical simplex, then one of the homology groups changes, so the two complexes have different homotopy types.
On the other hand, if the difference is an interval consisting of two or more simplices, then this interval defines a collapse, which preserves the homotopy type.
Since $f$ and $g$ have the same critical simplices---with the same values---the two complexes change in parallel so $\Alpha{r}{A}$ and $\Alpha{r}{\chi}$ have the same homotopy type for every $r$.

\smallskip
We note, however, that $f$ is not necessarily equal to $g$ restricted to the simplices in $\Delaunay{A}$.
Hence, the sequence of collapses and simplex deletions prescribed by $g$ does not necessarily restrict to collapses and deletions in $\Delaunay{A}$.
In other words, it is not clear whether or not the homotopy equivalence between corresponding sublevel sets of $f$ and $g$ can be strengthened to a simple homotopy equivalence:
\smallskip \begin{itemize}
  \item Is it true that $\Alpha{r}{\chi}$ collapses to $\Alpha{r}{A}$ for every $r \geq 0$?
\end{itemize}
Besides mathematical curiosity, a motivation to study this question is the possibility to speed up algorithms that compute properties of and relations between chromatic alpha complexes and their subcomplexes.
An effort in this direction can be found in \cite{NCBJ24}, where the authors generalize a result in \cite{BaEd17} to prove that the chromatic alpha filtration is related to the \v{C}ech filtration by simplicial collapses.

\section{Topological Summaries}
\label{sec:4}

When cancer cells are concentrated, is it clinically favorable if the immune cells are distributed around them or if they are evenly distributed within the tissue of interest? 
How to quantify the difference between these two, and possibly other classes of configurations is the motivation for this section. 
Mathematically speaking, the problem is about finding a \emph{sensible} quantitative measure of the spatial interactions between two or more point sets. 
The approach surveyed here is based on the theory of persistence for images, kernels, and cokernels developed in \cite{CEHM09}, which we introduce first.

\subsection{Algebraic Framework}
\label{sec:4.1}

Persistent homology, as introduced in Section~\ref{sec:2.2}, starts with a nested sequence of spaces, called a \emph{filtration}, and considers the corresponding sequence of homology groups and the maps between them. 
We study two parallel filtrations, with inclusion maps between them.
The extension proposed in \cite{CEHM09} considers the setting where we have a subspace $Y \subseteq X$, together with a function $f$ defined on $X$ and the goal is to track how the persistent homology of the space $Y$ relates to the persistent homology of the bigger space $X$. We call $g$ the restriction of $f$ to $Y$. 
The corresponding sequences of sublevel sets give rise to two parallel sequences of homology groups,
\smallskip \begin{center}
    \begin{tikzcd}
      \Hgroup{p}(X_0) \ar[r] & \Hgroup{p}(X_1) \ar[r] & \ldots \ar[r] & \Hgroup{p}(X_m) \\
     \Hgroup{p}(Y_0) \ar[r] \ar[u, "\kappa_0"] & \Hgroup{p}(Y_1) \ar[r] \ar[u, "\kappa_1"] & \cdots \ar[r] & \Hgroup{p}(Y_m) \ar[u, "\kappa_m"] 
    \end{tikzcd}
\end{center} \smallskip
for each dimension $p$, in which $X_i$ and $Y_i$ are the sublevel sets of $f$ and $g$ for the same value. 
The two sequences are connected by maps $\kappa_i \colon \Hgroup{p}(Y_i) \to \Hgroup{p}(X_i)$ induced by the inclusions $Y_i \subseteq X_i$. 
We are interested in the kernels, images, and cokernels of the connecting maps. 
In particular, it is easy to prove that each square in the diagram commutes, and that there are induced maps between consecutive kernels, images, and cokernels, respectively.
Homology classes are born and die in these sequences, just like in the sequences of homology groups.
We can therefore define persistent kernels, persistent images, and persistent cokernels as well as construct the corresponding persistence diagrams.
As proved in \cite{CEHM09}, these diagrams are stable, and there are fast algorithms to compute them.

\begin{figure}[hbt]
    \centering
    \includegraphics[width=.95\textwidth]{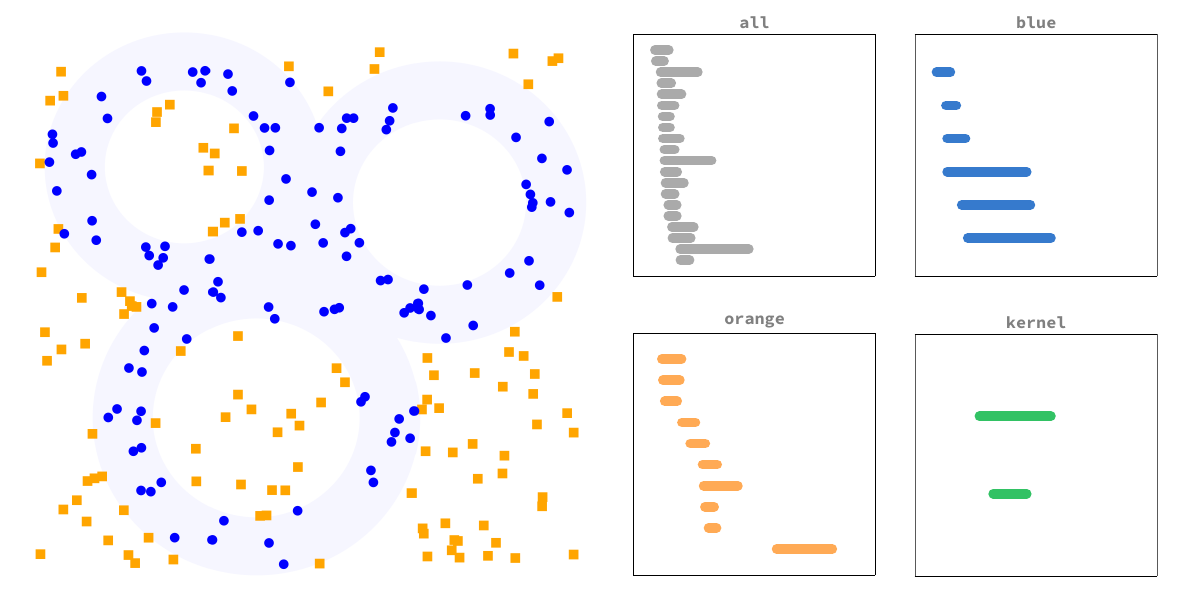}
    \caption{\footnotesize \emph{Left:} a set of \emph{blue} points sampled within the indicated three annuli, and another set of \emph{orange} points sampled outside these annuli so that one blue loop is filled, one half-filled and one empty.
    \emph{Right:} the (simplified) barcodes for all points, the \emph{blue} points, the \emph{orange} points, and for the kernel of the blue points included into the union.}
    \label{fig:monkey-set}
\end{figure}

\smallskip
As an example, consider the sets $A_0$ and $A_1$ of blue and orange points in the left panel of Figure~\ref{fig:monkey-set}.
To the right of that panel, we see four persistence barcodes---for the cycles counted by $1$-dimensional homology---each simplified by removing very short bars for better visibility.
The first three of these barcodes are for the radius functions on $\Delaunay{A_0 \cup A_1}$, $\Delaunay{A_0}$, and $\Delaunay{A_1}$, respectively.
They carry no information about the interaction between the two colors.
In the blue barcode, we see three long bars, which correspond to the holes of the three (shaded) annuli from which the blue points are sampled.
In the set of all points (blue and orange), two of these holes are partially filled by orange points, which is the reason why the corresponding (gray) bars are somewhat shorter than for the blue points alone.
In the orange barcode, the longest bar is born relatively late, because the corresponding hole is only partially surrounded by orange points.
The message we want to convey is that in contrast to these three, the kernel barcode speaks directly about the interaction between blue and orange points.
To construct it, let $L = \Delaunay{A_0}$ and $K = \Delaunay{\chi}$ represent the blue points and all points, respectively, and consider the kernels of the maps $\kappa_i \colon L_i \to K_i$ induced by the inclusions.
There are only two long bars, which correspond to the holes in the blue set that are partially filled by orange points.
The unfilled third hole remains in the image of $\kappa_i$ throughout its existence and thus makes no appearance in the kernel.

\subsection{6-pack of Persistent Diagrams}
\label{sec:4.2}

The main concept in this section is a collection of six related persistence diagrams, which we use to quantify the way point sets mingle. 
We call it a \emph{$6$-pack}, which can be defined for any pair of topological spaces $L \subseteq K$ with a filtration on $K$.
For example, $K$ may be the chromatic Delaunay mosaic of the points in Figure~\ref{fig:monkey-set} and $L$ may be the Delaunay mosaic of just the blue points, as discussed above.
Let $\Kfun \colon K \to \Rspace$ be the chromatic radius function, and write $\Lfun$ and $\KLfun$ for the restrictions of $\Kfun$ to $L$ and $K \setminus L$.
The three radius functions are used to compute three persistence diagrams: for $K$, $L \subseteq K$, and the pair, $(K,L)$.
Indeed, for $L$, we reduce its boundary matrix ordered by $\Lfun$, while for $(K,L)$, we get the diagram by reducing the ordered boundary matrix of $K$ after purging all rows and columns of simplices in $L$.
We get three additional diagrams for the kernel, image, and cokernel of the map on homology induced by the inclusion $L \subseteq K$.
The diagrams in the $6$-pack are arranged as in Table~\ref{tab:6-pack}, in a manner that lends itself to comparing the information between them.
\begin{table}[htb]
  \centering
  \begin{tabular}{l||l||l}
    \emph{kernel:} & \emph{relative:} & \emph{cokernel:} \\
    $\Dgm{}{\kernel{}{\Lfun \to \Kfun}}$ & $\Dgm{}{\KLfun}$ & $\Dgm{}{\coker{}{\Lfun \to \Kfun}}$ \\ \hline \hline
    \emph{domain:} & \emph{image:} & \emph{codomain:} \\
    $\Dgm{}{\Lfun}$ & $\Dgm{}{\image{}{\Lfun \to \Kfun}}$ & $\Dgm{}{\Kfun}$ \\ [+2mm]
  \end{tabular}
  \caption{\footnotesize The arrangement of the persistence diagrams in the $6$-pack for the pair $L \subseteq K$ in two rows and three columns.
  Read the six positions in a circle so that the domain lies between the kernel and the image, the image lies between the domain and the codomain, etc.}
  \label{tab:6-pack}
\end{table}

\begin{figure}[hbt]
    \centering
    \includegraphics[width=.97\textwidth]{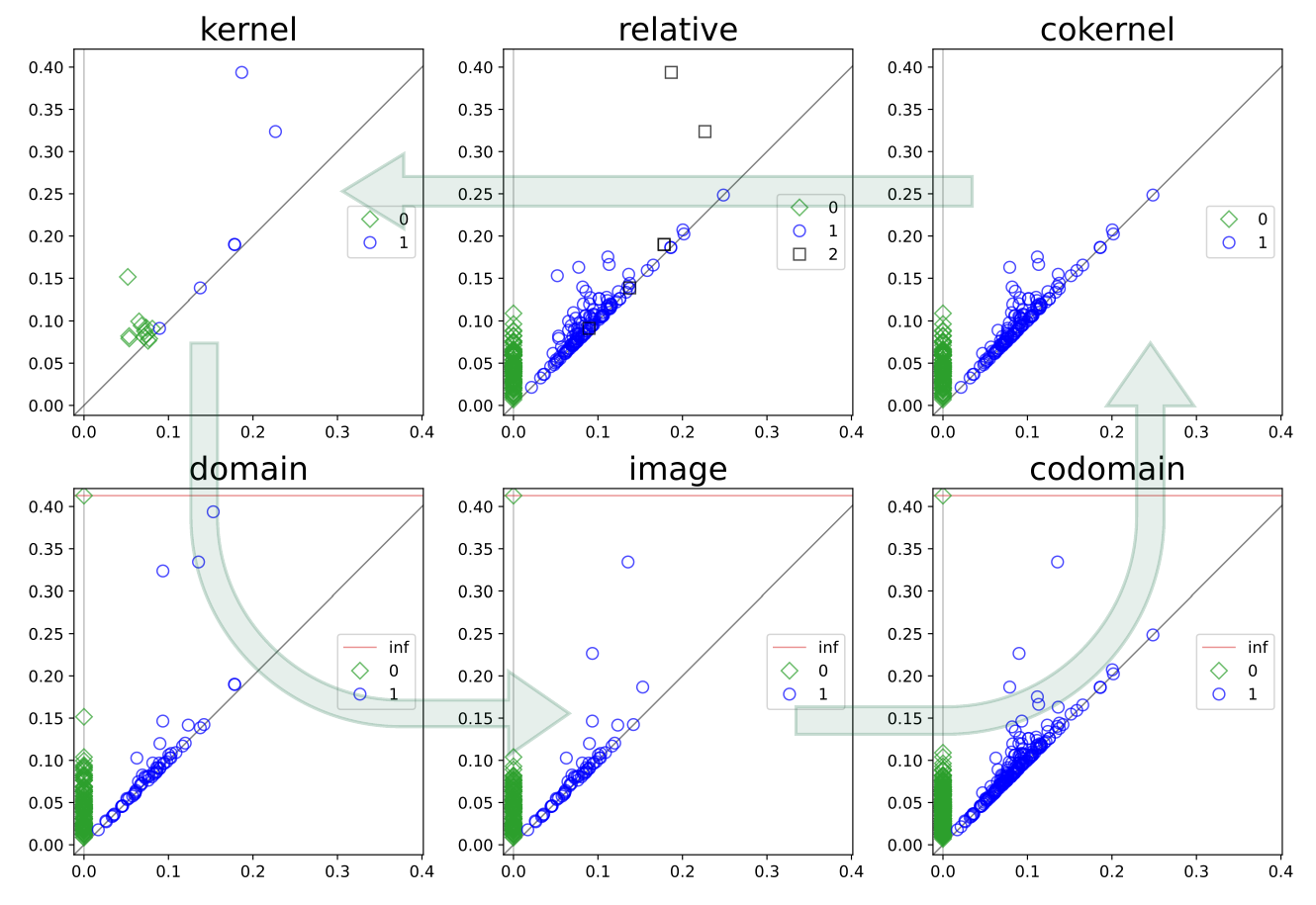}
    \caption{\footnotesize The $6$-pack of persistence diagrams for the inclusion of the blue complex into the chromatic complex for the points in Figure~\ref{fig:monkey-set}.
    The three arrows indicate the short exact sequences that give rise to the relations between the $1$-norms stated in Theorem~\ref{thm:norm_relations}.}
    \label{fig:six-pack}
\end{figure}

\smallskip
We illustrate the concepts with the $6$-pack for the $2$-colored set in Figure~\ref{fig:monkey-set}, which we show in Figure~\ref{fig:six-pack}. 
The domain diagram contains three persistent loops, as expected. 
One of these loops is not filled by orange points, which implies that its corresponding point in the domain diagram also appears in the image diagram. 
The other two persistent blue loops contain orange points inside, leading to two points relatively high above the diagonal in the kernel diagram.
Note that the relative diagram has points representing $2$-dimensional relative homology classes at the same locations.
This is because the fillings that give death to the cycle appear as relative $2$-cycles.
The interested reader is encouraged to try the software in \cite{DrMa23}, which also supports $3$-dimensional point sets.
It is based on the matrix reduction algorithms for computing kernel, image, and cokernel persistence described in \cite{CEHM09}.

\subsection{Properties of the 6-pack}
\label{sec:4.3}

Clearly, the information contained in any one persistent diagram is not independent of the information in the other five diagrams in the $6$-pack.
To mention one example: a class in the image dies either because it also dies in the domain, or it is born in the kernel.
This particular relation has an algebraic expression as an exact sequence involving the kernel, the domain, and the image.
There are two additional exact sequences of the map $\kappa_i \colon \Hgroup{p}(L_i) \to \Hgroup{p}(K_i)$ induced by the inclusion $L_i \subseteq K_i$:
\begin{align}
    0 &\to \kernel{p}{\kappa_i} \to \Hgroup{p}(L_i) \to \image{p}{\kappa_i} \to 0;
    \label{eqn:ranksg} \\
    0 &\to \image{p}{\kappa_i} \to \Hgroup{p}(K_i) \to \coker{p}{\kappa_i} \to 0;
    \label{eqn:ranksf} \\
    0 &\to \coker{p}{\kappa_i} \to \Hgroup{p}(K_i,L_i) \to \kernel{p-1}{\kappa_i} \to 0,
    \label{eqn:ranksm}
\end{align}
for each dimension $p$; see Figure~\ref{fig:six-pack} where the exact sequences are indicated by arrows that traverse the diagrams in a cyclic order.
As a consequence of these sequences, we get relations between the sums of persistences of the points in the six diagrams.
Define the \emph{$1$-norm} of the $p$-dimensional persistence diagram of $f_K \colon K \to \Rspace$ as the total difference between the deaths and the births in this diagram, denoted $\norm{\Dgm{p}{\Kfun}}_1$, and similarly for $f_L$ and $f_{K,L}$.
Some of the classes may be born but never die, which would render this definition meaningless.
To finesse this difficulty, we use a threshold, $C$, larger than all births, and let every class die at $C$ unless it has died earlier.
With this caveat, we can prove the following theorem.
\begin{theorem}\cite[Theorem 5.3]{CDES22}
  \label{thm:norm_relations}
  Let $L \subseteq K$ be simplicial complexes, $\Kfun \colon K \to \Rspace$ monotonic, and $\Lfun$, $\KLfun$ the restrictions of $\Kfun$ to $L$ and $K \setminus L$.
  For each dimension, $p$, we have
  \begin{align}
    \norm{\Dgm{p}{\Lfun}}_1  &=  \norm{\Dgm{p}{\kernel{}{\Lfun \to \Kfun}}}_1 + \norm{\Dgm{p}{\image{}{\Lfun \to \Kfun}}}_1 ,
      \label{eqn:onenormg} \\
    \norm{\Dgm{p}{\Kfun}}_1  &=  \norm{\Dgm{p}{\image{}{\Lfun \to \Kfun}}}_1 + \norm{\Dgm{p}{\coker{}{\Lfun \to \Kfun}}}_1 , 
      \label{eqn:onenormf} \\
    \norm{\Dgm{p}{\KLfun}}_1  &=  \norm{\Dgm{p}{\coker{}{\Lfun \to \Kfun}}}_1 + \norm{\Dgm{p-1}{\kernel{}{\Lfun \to \Kfun}}}_1 .
      \label{eqn:onenormm}
  \end{align}
\end{theorem}
Despite these relations between the diagrams in a $6$-pack, no single diagram is necessarily determined by the five other diagrams; see \cite[Figure~13]{CDES22}.

\subsection{Choosing the Subcomplex}
\label{sec:4.4}

What other choices do we have apart from $L$ corresponding to one color and $K$ corresponding to all colors? 
How can we use three or more colors in a way that is not a mere reduction to the two color case? 
In \cite{CDES22}, the \emph{$t$-chromatic subcomplex} of a chromatic alpha complex is defined as the collection of those simplices whose vertices have $t$ colors or fewer.
For example, the mono-chromatic subcomplex of a chromatic Delaunay complex is the disjoint union of the Delaunay complexes of each color.
Similarly, the bi-chromatic subcomplex contains all simplices connecting vertices of at most two colors, but does not contain simplices connecting vertices of three or more colors.

\smallskip
We can now chose different combinations of subcomplexes for $L$ and $K$. Consider the kernel persistence diagram as in Figure~\ref{fig:monkey-set} and some point set with three colors. If we choose the inclusion of the mono-chromatic subcomplex into the bi-chromatic subcomplex, we will detect loops created by any one color that are filled by any one of the other colors. 
On the other hand, if we instead choose the inclusion of the bi-chromatic subcomplex into the full tri-chromatic complex, we detect loops created by any two colors, that are filled by the third color.

\section{Extracting Mingling Numbers}
\label{sec:5}

The wealth of information contained within the 6-pack of diagrams can be challenging to comprehend. 
This section explores whether there are meaningful and interpretable numbers that can be extracted from these diagrams. 
Our aim is to develop a numerical summary that quantifies the extent to which two or more point sets are mixed or mingle.

\subsection{From Shares to the MST-ratio}
\label{sec:5.1}

Consider a $2$-colored point set, and the inclusion of the mono-chromatic subcomplex into the full bi-chromatic alpha complex.
When points of different colors are well-separated---say by a sufficiently wide corridor---then the kernel diagram is necessarily empty.
By \eqref{eqn:onenormg} in Theorem~\ref{thm:norm_relations}, the $1$-norms of the kernel and the image diagrams add up to the $1$-norm of the domain diagram.
Since the kernel is empty, this implies that every homology class in the domain is also in the image.
But as we allow one color to invade the area of the other, homology classes will migrate from the image to the kernel.
In other words, the kernel and the image \emph{share} the classes in the domain, which suggests we use the share of either as a measure of how mixed the two colors are.

\smallskip
Following this suggestion, we relate these shares to intuitive geometric concepts in the setting of $0$-dimensional homology for a bi-chromatic set in $\Rspace^2$.
To explain, let $A \subseteq \Rspace^2$ be finite, write $A_r$ for the union of disks of radius $r \geq 0$ centered at the points in $A$, and note the close relation between the $0$-dimensional homology of $A_r$ and the (Euclidean) minimum spanning tree of $A$, denoted $\MST{A}$.
Specifically, the connected components of $A_r$ correspond to the connected subtrees we obtain by removing all edges longer than $2r$ from $\MST{A}$.
It follows that the $1$-norm of the $0$-dimensional persistence diagram of the chromatic alpha complexes is half the length of $\MST{A}$.
Write $B \subseteq A$ and $C = A \setminus B$ for the points of one color and the other.
We consider the map from the mono-chromatic to the bi-chromatic complexes.
In other words, we consider the disjoint union of the two sets of disks, $B_r \sqcup C_r$, and the map induced by the inclusions $B_r \subseteq A_r$ and $C_r \subseteq A_r$.
Ignoring the classes that never die, the $1$-norm of the persistence diagram of the domain of this map is half the combined lengths of the two minimum spanning trees, $\frac{1}{2} \Length{\MST{B}} + \frac{1}{2} \Length{\MST{C}}$,
and it is not difficult to see that the $1$-norm of the image persistence diagram is $\frac{1}{2} \Length{\MST{A}}$.
Formulated as fractions, the \emph{image share} is
${\Length{\MST{A}}}/\left({\Length{\MST{B}}+\Length{\MST{C}}}\right)$, and the \emph{kernel share} is $1$ minus the image share.
We prefer the reciprocals and define the \emph{MST-ratio} as one over the image share:
\begin{align}
  \text{MST-ratio} &= \frac{\Length{\MST{B}}+\Length{\MST{C}}}{\Length{\MST{A}}} .
\end{align}
The union of $\MST{B}$ and $\MST{C}$ connected by one additional edge is a spanning tree that is by definition at least as long as $\MST{A}$. 
While the ratio can be smaller than $1.0$, it is generally larger than that, and the larger it is, the more mixed the sets $B$ and $C$ appear.

\subsection{Bounds for the MST-ratio}
\label{sec:5.2}

To what extent can the minimum spanning trees of two finite sets be longer than the one minimum spanning tree of the union of the two sets?
Given this union, we are interested in the maximum ratio, over all partitions into two sets, and, in particular, in the infimum and supremum of this maximum, over all sets in a given class. 
Here we consider four classes of point sets in $\Rspace^2$: all finite sets, points drawn uniformly at random, dense sets with specified spread, and lattices.
For a given set $A \subseteq \Rspace^2$, we write $\mu (A)$ for the maximum MST-ratio, over all subsets $B \subseteq A$.
For random sets, we are interested in the expected maximum MST-ratio, while for the other three classes we study the infimum, $\inf \mu (A)$, and the supremum, $\sup \mu (A)$, over all sets $A$ is the class at hand.
\begin{theorem}
  \label{thm:bounds_for_MST-ratio}
  Let $A$ be a set of $n$ points in $\Rspace^2$.
  \begin{enumerate}
    \item[(i)] For the class of all finite sets, we have $\inf \mu (A) = 1$ and $2.154 \leq \sup \mu (A) \leq 2.427$.
    \item[(ii)] For the class of uniformly random points in $[0,1]^2$, the expected MST-ratio tends to $\sqrt{2}$ as $n$ goes to infinity.
    \item[(iii)] For the class of sets with spread at most some constant times $\sqrt{n}$, $\inf \mu (A) > 1$.
    \item[(iv)] For lattices, $\inf \mu (A) = 1.25$ and $\sup \mu (A) = 2$.
  \end{enumerate}
\end{theorem}
We briefly discuss each of the four results.
It is not difficult to see that there are sets in $\Rspace^2$ for which the two trees can be barely larger than the one tree, for all possible bipartitions.
By comparison, the sup-max is more interesting.
It relates to the classic Steiner tree question: how much shorter can we make the minimum spanning tree of a set if we are allowed to add points to the set?
The conjectured supremum ratio is ${2}/{\sqrt{3}} = 1.154\ldots$ \cite{GiPo68}, but the current best upper bound is only $1.213\ldots$ \cite{ChGr85}.
Both bounds for the sup-max are derived from this bound on the Steiner ratio \cite{CDES24}.

\smallskip
The average MST-ratio is a lower bound for the maximum MST-ratio. 
For $n$ points sampled uniformly at random in $[0,1]^2$, the maximum MST-ratio is at least $\sqrt{2} - \ee$, for every $\ee > 0$, with probability tending to $1$ as $n$ goes to infinity; see \cite{DPT23} where this lower bound proved using classic work on the expected length of the minimum spanning tree by Beardwood, Halton and Hammersley~\cite{BHH59}.
A tighter analysis shows that the expected average MST-ratio tends to $\sqrt{2}$ as $n$ goes to infinity \cite{DrSa24}.

\smallskip
The \emph{spread} of $n$ points is the diameter divided by the minimum inter-point distance.
In $\Rspace^2$, the spread is at least some constant times $\sqrt{n}$.
A lower bound for $\inf \mu (A)$ that is strictly larger than $1$ but depends on this constant can be found in \cite{DPT23}.
The proof is based on partitions that include lattice-like subsets, so the results for lattices proved in \cite{CDES24} are relevant.
Since a lattice in $\Rspace^2$ is necessarily infinite, we consider progressively larger finite portions---each the intersection with a disk or square centered at the origin---and take the limit of the MST-ratios.
As proved in \cite{CDES24}, the maximizing bi-partition for the hexagonal lattice has MST-ratio $1.25$, and every other lattice achieves $1.25$ or higher.
Furthermore, there is a lattice with maximum MST-ratio approaching $2$, and no other lattice approaches a higher ratio.

\subsection{Generalizations}
\label{sec:5.3}

Above we considered the MST-ratio or, equivalently, the image share for $s+1=2$ colors, points in dimension $d=2$, and homology dimension $p=0$.
We sketch five potential avenues for further research to broaden but also deepen the study:
\begin{enumerate}
  \item[(i)] three or more colors $(s+1 \geq 3)$;
  \item[(ii)] points in three or more dimensions $(d \geq 3)$;
  \item[(iii)] homology beyond zero dimension $(p \geq 1)$;
  \item[(iv)] ratios different from the kernel and image shares;
  \item[(v)] maps different from the mono-chromatic to the bi-chromatic complex.
\end{enumerate}
We formulate specific questions for the first three research directions.
Considering the case $s+1 = 3$, $d = 2$, $p = 0$, let $A \subseteq \Rspace^2$ be the hexagonal lattice, and $\chi \colon A \to \{0,1,2\}$ a $3$-coloring or, equivalently, a partition into three sets.
If the three colors are hexagonal sublattices isomorphic to each other, then the MST-ratio---defined as the combined length of the three mono-chromatic MSTs over the length of the tri-chromatic MST---approaches $\sqrt{3}$ as we take progressively larger finite portions of $A$.
\begin{itemize}
  \item Is it true that $\sqrt{3}$ is the supremum MST-ratio over all $3$-colorings of the hexagonal lattice?
  \item Is $\sqrt{3}$ the infimum, over all lattices, of the maximum MST-ratio, over all $3$-colorings?
\end{itemize}
Considering the case $s+1 = 2$, $d = 3$, $p = 0$, let $A = {\rm FCC}$ be the face-centered cubic lattice, which consists of all integer points whose sums of coordinates are even, let $B = 2{\rm FCC}$ be the sublattice with twice the inter-point distances, and let $C = A \setminus B$ be the rest.
Observe that the MST-ratio of this bi-coloring converges to $\sfrac{9}{8} = 1.125$.
\begin{itemize}
  \item Is it true that no $2$-coloring of the FCC lattice achieves a larger MST-ratio?
  \item Is $\sfrac{9}{8}$ the infimum, over all lattices in $\Rspace^3$, if the maximum MST-ratio, over all $2$-colorings?
\end{itemize}
Considering the case $s+1 = 2$, $d = 2$, $p = 1$, let $A \subseteq \Rspace^2$ be the integer lattice, and let $\chi \colon A \to \{0, 1\}$ be the \emph{checkerboard coloring}, in which $\chi(a)$ is the parity of the sum of the two coordinates of $a$.
Mapping the mono-chromatic complex to the bi-chromatic complex, every cycle born in the domain dies at the same radius in the codomain, so the $1$-dimensional persistence diagram of the image is empty.
By \eqref{eqn:onenormg} in Theorem~\ref{thm:norm_relations}, this implies that the $1$-dimensional persistence diagram of the kernel is that of the codomain.
Each cycle is born at $r = \sfrac{1}{2}$ and dies at $r = \sfrac{\sqrt{2}}{2}$, which implies that the $1$-norm converges to $\frac{1}{2} ( \sqrt{2} - 1 ) = 0.207\ldots$ times the number of unit squares in the finite portion of the lattice considered.
\begin{itemize}
  \item Is there a geometric rephrasing of the kernel and image shares for ($1$-dimensional) cycles that is similarly compelling as the MST-ratio for $0$-dimensional homology?
\end{itemize}
The MST-ratio is motivated by the short exact sequence \eqref{eqn:ranksg}.
We can similarly define ratios following the sequences \eqref{eqn:ranksf} and \eqref{eqn:ranksm}.
There are, however, more than two options---even for two colors in $\Rspace^2$---because there is more than one map that may be explored with the corresponding $6$-pack.
Finally, we mention that currently we do not know whether or not there is a polynomial-time algorithm for coloring the points to maximize the MST-ratio even just for two colors.

\section{Discussion}
\label{sec:6}

Biological inquiries have been a driving force in the development of geometric and topological methodologies. 
The introduction of 3D alpha shapes in the 1990s to capture the structures of biomolecules from the positions of their atoms exemplifies this trend.
Depending on the radius parameter, the alpha shape reflects structural motifs on different scales and combines them to a continuous hierarchy of representations.
Similarly, the chromatic alpha complexes arose from the biomedical need to analyze complex spatial interactions among multiple cell types within systems like tumor microenvironments. 
By color-coding different cell types, we obtain stable topological summaries that quantify inter-cellular interactions at all scales.
Just as the alpha shape provides a quantitative description of the shape of a biomolecule, the chromatic extension offers a quantitative language for cellular interaction.

\smallskip
Notwithstanding the biological applications, the chromatic alpha complexes reveal a rich geometric and topological structure that is compelling in its own right.
This survey reviews what is known at the time of writing, and these early results suggest a number of directions for further research beyond the more specific questions stated in the main text of this survey.
Similar to the setting without colors, the extension from unweighted to weighted points is straightforward while significantly expanding the range of possible applications, e.g.\ to represent different atom types in materials.
More challenging is basing the construction of the chromatic alpha complexes and the corresponding $6$-packs on order-$k$ rather than the ordinary order-$1$ Delaunay mosaics.
Another important direction is the stochastic analysis of chromatic persistent homology, which can serve as a baseline for comparison, helping to identify the non-random features of data.
The algorithms are sufficiently fast to approach this question experimentally, e.g.\ by computing norms of the diagrams in the $6$-packs for random colorings and ratios between them.



\begin{thebibliography}{21}

\footnotesize{

\bibitem{AENY23}
{\sc H.\ Asashiba, E.G.\ Escolar, K.\ Nakashima and M.\ Yoshiwaki.}
On approximation of 2D persistence modules by interval-decomposables.
\texttt{arXiv:1911.01637 [math.RT]}, 2023.

\bibitem{BaEd17}
{\sc U.\ Bauer and H.\ Edelsbrunner.}
The Morse theory of \v{C}ech and Delaunay complexes.
\emph{Trans.\ Amer.\ Math.\ Soc.} {\bf 369} (2017), 3741--3762.


\bibitem{BHH59}
{\sc J.\ Beardwood, J.H.\ Halton and J.M.\ Hammersley.}
The shortest path through many points.
\emph{Math.\ Proc.\ Cambridge Phil.\ Soc.} {\bf 55} (1959), 299--327.

\bibitem{BCKO08}
{\sc M.\ de Berg, O.\ Cheong, M.\ van Kreveld and M.\ Overmars.}
\emph{Computational Geometry: Algorithms and Applications.}
Third edition, Springer-Verlag, Berlin, Heidelberg, 2008.


\bibitem{BCDES22}
{\sc R.\ Biswas, S.\ Cultrera di Montesano, O.\ Draganov, H.\ Edelsbrunner and M.\ Saghafian.}
On the size of chromatic Delaunay mosaics. 
\texttt{arXiv:2212.03121 [math.CO]}, 2022.



\bibitem{Car09}
{\sc G.\ Carlsson.}
Topology and data.
\emph{Bull.\ Amer.\ Math.\ Soc.} {\bf 46} (2009), 255--308.

\bibitem{ClSh89}
{\sc K.L.\ Clarkson and P.\ Shor.}
Applications of random sampling.
\emph{Discrete Comput.\ Geom.} {\bf 4} (1989), 387--421.

\bibitem{CEH07}
{\sc D.\ Cohen-Steiner, H.\ Edelsbrunner and J.\ Harer.}
Stability of persistence diagrams.
\emph{Discrete Comput.\ Geom.} {\bf 37} (2007), 103--120.

\bibitem{CEHM09}
{\sc D.\ Cohen-Steiner, H.\ Edelsbrunner, J.\ Harer and D.\ Morozov.}
Persistent homology for kernels, images, and cokernels.
In ``Proc.\ 20th Ann.\ ACM-SIAM Sympos.\ Discrete Alg., 2009'', 1011--1020.

\bibitem{ChGr85}
{\sc F.R.K.\ Chung and R.L.\ Graham.}
A new bound for Euclidean Steiner minimal trees.
In \emph{Discrete Geometry and Convexity}, Annals N.Y.\ Acad.\ Sci.\ {\bf 440}, New York, 1985, 328--346.


\bibitem{CDES22}
{\sc S.\ Cultrera di Montesano, O.\ Draganov, H.\ Edelsbrunner, and M.\ Saghafian.} 
Chromatic alpha complexes. \texttt{arXiv:2212.03128} (2024). 

\bibitem{CDES24}
{\sc S.\ Cultrera di Montesano, O.\ Draganov, H.\ Edelsbrunner, and M.\ Saghafian.} 
The Euclidean MST-ratio for 2-colored Lattices.
\texttt{arXiv:2403.10204} (2024).

\bibitem{Del34}
{\sc B.\ Delaunay.}
Sur la sph\`{e}re vide.
\emph{Izv.\ Akad.\ Nauk SSSR, Otdelenie Matematicheskii i Estestvennykh Nauk} {\bf 7} (1934), 793--800. 

\bibitem{DrMa23}
{\sc O.\ Draganov and M.\ Mahini.}
Chromatic-tda.
\texttt{github.com/OnDraganov/chromatic-tda}, 2023.

\bibitem{DrSa24}
{\sc O.\ Draganov and M.\ Saghafian.}
Private communication, 2024.

\bibitem{DPT23}
{\sc A.\ Dumitrescu, J.\ Pach and G.\ T\'{o}th.}
Two trees are better than one.
\texttt{arXiv:2312.09916 [math.CO]}, 2023.

\bibitem{EdHa10}
{\sc H.\ Edelsbrunner and J.L.\ Harer.}
\emph{Computational Topology. An Introduction.}
Amer.\ Math.\ Soc., Providence, Rhode Island, 2010.


\bibitem{EdKiSe83}
{\sc H.\ Edelsbrunner, D. G.\ Kirkpatrick and R.\ Seidel.}
On the shape of a set of points in the plane. 
\emph{IEEE Trans.\ Inform.\ Theory} {\bf IT-29} (1983), 551-–559.

\bibitem{EdMu94}
{\sc H.\ Edelsbrunner and E.P.\ M\"{u}cke.}
Three-dimensional alpha shapes.
\emph{ACM Trans.\ Graphics} {\bf 13} (1994), 43--72

\bibitem{For98}
{\sc R.\ Forman.}
Morse theory for cell complexes.
\emph{Adv.\ Math.} {\bf 134} (1998), 90--145.

\bibitem{GiPo68}
{\sc E.N.\ Gilbert and H.O.\ Pollak.}
Steiner minimal trees.
\emph{SIAM J.\ Appl.\ Math.} {\bf 16} (1968), 1--29.


\bibitem{Hat02}
{\sc A.\ Hatcher.}
\emph{Algebraic Topology.}
Cambridge Univ.\ Press, Cambridge, England, 2002.


\bibitem{Kru56}
{\sc J.B.\ Kruskal.}
On the shortest spanning tree of a graph and the traveling salesman problem.
\emph{Prof.\ Amer.\ Math.\ Soc.} {\bf 7} (1956), 48--50.





\bibitem{NCBJ24}
{\sc A.\ Natarajan, T.\ Chaplin, A.\ Brown and M.-J.\ Jimenez.}
Morse theory for chromatic Delaunay triangulations.
\texttt{arXiv:2405.19303 [math.AT]}, 2024.

\bibitem{PFRT22}
{\sc G.\ Palla, D. S.\ Fischer, A.\ Regev and F. J.\ Theis.}
Spatial components of molecular tissue biology.
\emph{Nat. Biotechnol.} {\bf 40} (2022), 308--318.

\bibitem{ReBo23}
{\sc Y.\ Reani and O.\ Bobrowski.}
A coupled alpha complex.
\emph{J.\ Comput.\ Geom.} {\bf 14} (2023), 221--256.

\bibitem{Sto23}
{\sc B.J.\ Stolz, J.\ Dhesi, J.A.\ Bull, H.A.\ Harrington, H.M.\ Byrne and I.H.R.\ Yoon.}
Relational persistent homology for multispecies data with application to the tumor microenvironment.
\texttt{arXiv:2308.06205 [math.AT]}, 2023.

\bibitem{Vor070809}
{\sc G.\ Voronoi.}
Nouvelles applications des param\`{e}tres continus \`{a} la th\'{e}orie des formes quadratiques.
\emph{J.\ Reine Angew.\ Math.} {\bf 133} (1907), 97--178 and {\bf 134} (1908), 198--287 and {\bf 136} (1909), 67--182.

\bibitem{WAWB24}
{\sc H.\ Wagner, N.\ Arustamyan, M.\ Wheeler and P.\ Bubenik.}
Mixup barcodes: quantifying geometric-topological interactions between point clouds.
\texttt{arXiv:2402.15058 [math.AT]}, 2024.

\bibitem{Zhi15}
{\sc B.\ Zhilinskii.}
\emph{Introduction to Louis Michel's Lattice Geometry through Group Action.}
EDP Sciences, ENRS Editions, Paris, France, 2015.




}

\end{thebibliography}
\end{document}